\documentclass [12pt, letterpaper]{article}
\usepackage{fullpage}
\usepackage{amsmath}
\usepackage{amssymb}
\usepackage{graphicx}
\usepackage{mathrsfs}
\usepackage{wrapfig}
\usepackage{boxedminipage}
\usepackage{epsfig}
\usepackage{setspace}
\usepackage{subfigure}
\usepackage{float}
\usepackage{hyperref}
\newcommand{\reef}[1]{(\ref{#1})}
\begin{document}

\begin{flushright}
\phantom{{\tt arXiv:0709.????}}
\end{flushright}

\bigskip
\bigskip
\bigskip

\begin{center} {\Large \bf Landau Levels, Magnetic Fields }
  
  \bigskip

{\Large\bf  and }

\bigskip

{\Large\bf Holographic   Fermi
    Liquids}

\end{center}

\bigskip \bigskip \bigskip \bigskip

\centerline{\bf Tameem Albash, Clifford V. Johnson}

\bigskip
\bigskip
\bigskip

  \centerline{\it Department of Physics and Astronomy }
\centerline{\it University of
Southern California}
\centerline{\it Los Angeles, CA 90089-0484, U.S.A.}

\bigskip

\centerline{\small \tt talbash,  johnson1,  [at] usc.edu}

\bigskip
\bigskip


\begin{abstract} 
\noindent 
We further consider a probe fermion in a dyonic black hole background
in anti--de Sitter spacetime, at zero temperature, comparing and
contrasting two distinct classes of solution that have previously
appeared in the literature. Each class has members labeled by an
integer $n$, corresponding to the $n$th Landau level for the
fermion. Our interest is the study of the spectral function of the
fermion, interpreting poles in it as indicative of quasiparticles
associated with the edge of a Fermi surface in the holographically
dual strongly coupled theory in a background magnetic field ${\cal H}$
at finite chemical potential. Using both analytical and numerical methods, we explicitly
show how one class of solutions naturally leads to an infinite family
of quasiparticle peaks, signaling the presence of a Fermi surface for
each level~$n$. We present some of the properties of these peaks,
which fall into a well behaved pattern at large $n$, extracting the
scaling of Fermi energy with $n$ and ${\cal H}$, as well as the
dispersion of the quasiparticles.
\end{abstract}
\newpage \baselineskip=18pt \setcounter{footnote}{0}


\section{Introduction}
%
Recent work has considerably enriched our detailed knowledge of the
kinds of strongly coupled physics that can be studied using
holographic studies of charged black holes.  Such studies are in fact
quite old, since it was realized very shortly after the formulation of
the AdS/CFT correspondence \cite{Maldacena:1997re,
  Witten:1998qj,Gubser:1998bc} that electrically charged black holes
allow for the study of non--trivial strongly coupled physics at finite
density or chemical potential\cite{Chamblin:1999tk,Chamblin:1999hg} at
both non--zero and zero temperature.  A rich phase structure was
uncovered in refs.\cite{Chamblin:1999tk,Chamblin:1999hg}, for (finite
volume) systems dual to asymptotically AdS spacetime backgrounds with
Reissner--Nordstr\"om black holes in diverse dimensions. (See also the
related work in refs.\cite{Cvetic:1999ne,Cvetic:1999rb} on the phase
structure of certain scalar charged black holes.)

A decade after those studies a new wave of interest emerged,
principally due to key observations about how charged black hole
solutions to asymptotically AdS Einstein--Maxwell systems (sometimes
coupled to scalars and fermions) may pertain directly to holographic
models of superconductivity
\cite{Gubser:2008px,Hartnoll:2008kx,Hartnoll:2008vx} and to
holographic access to Fermi
surfaces\cite{Liu:2009dm,Faulkner:2009wj}. (Note that the recent focus
has been infinite volume system, therefore focusing on flat black
holes.)

In the latter case (refs.\cite{Liu:2009dm,Faulkner:2009wj}) the system
that displays Fermi surface behaviour is a zero temperature
Reissner--Nordstr\"om black hole with a probe fermion.  The spectral
function $G_R(\omega, k)$ of the probe fermion in this background was
shown, at $T = 0$ and finite chemical potential $\mu$, to exhibit a
sharp peak at $\omega = 0$ and finite $k ( = k_F)$. This peak has been
interpreted as the quasiparticle pole at the edge of a Fermi surface.
Here $\omega$ is identified with Fermi energy above the chemical
potential, $\omega = E_F - \mu$. The dispersion associated with this
peak was observed to be distinctly non--Landau in a manner consistent
with the quasiparticles being not effectively free (as in standard
Landau theory), but in a strongly coupled phase. This is encouraging,
since this type of behaviour is reminiscent of that which one might
hope to gain understanding of in order to gain insight into various
novel phenomena known from experiment in condensed matter physics (see
for example ref.\cite{varma-2002-361} for a review).

Several concerns arise here.  Not the least is that the while the
results are suggestive, a quasiparticle peak alone (especially for a
minimally coupled non--backreacting fermion) is not damning evidence
for a Fermi surface.  Some encouragement is to be found in
ref.~\cite{Cubrovic:2009ye} where they showed that at $T \neq 0$ the
peak broadened as a function of $T$ in the expected manner, but much
more evidence is desirable.  Another concern is that the parallel
studies of holographic superconductivity in the literature all
strongly indicate that a charged scalar profile (representing the
condensate/order parameter) would have developed at a critical
temperature well above $T = 0$, and so an extremal
Reissner--Nordstr\"om solution is presumably the wrong object to be
studying for the $T=0$ physics at $\mu \neq 0$.  (In fact, that an
extremal black hole is probably excluded from this point of the phase
diagram was already suggested a decade ago\cite{Chamblin:1999tk},
since there is an expected decay channel due to super--radiance
effects\cite{Penrose:1969pc}, inherited from the fact that the charged
black holes are really dimensional reductions of spinning D3--branes.)
The approximate holographic superconductivity studied in
refs.~\cite{Gubser:2008px,Hartnoll:2008kx,Hartnoll:2008vx} suggest
that at $T=0$ the system would be (at best) a scalar--profiled charged
black hole (\emph{i.e.} not Reissner--Nordstr\"om) and so one might
argue that this is the system to which the spectral analysis should be
applied.  However, further study in a fully backreacted M--theory
context\cite{Gauntlett:2009dn,Gubser:2009qm,Gubser:2009gp,Gauntlett:2009bh}
has shown that the black hole nature of the system entirely vanishes
at $T=0$.  An AdS background re--emerges at $T=0$ instead.  This may
all fit well with the expectation that if the system has become
superconducting below some $T=T_c$ the Fermi surface may well no
longer be available, although it remains to be seen what the
appropriate study might reveal.

With all these concerns up front, it is our view that the work of
refs.~\cite{Liu:2009dm,Faulkner:2009wj}
remains of considerable value.  It is a simple, remarkable system with
many of the key features of interest for some of the studies, since
away from $T=0$, at high enough temperature, a charged black hole will
 still be relevant.  The techniques developed and insights gained in
this study, even if the system is not exactly the right one, will be
of value in the broader context.

In this spirit, we continue the study that we began of the
Reissner--Nordstr\"om system and the putative Fermi surface in the
presence of a background magnetic field ${\cal H}$.  In our first
paper\cite{Albash:2009wz}, we studied a dyonic black hole, the
magnetic component now playing the role of an external magnetic field
of the system.  On general grounds, one expects the system to develop
an infinite set of Landau levels, the quantized energy levels of the
fermions in the magnetic field.  Our first paper studied the lowest
Landau level, and already we saw interesting physics, including
multiple quasiparticle peaks and a magnetic field dependence in the
dispersion characteristic of the peaks.

The non--zero magnetic field case has peaks appearing at $\omega = E_F
- \mu \neq 0$ since now the energy of the fermions has been increased
as compared to the ${\cal H}= 0$ system, for a given~$\mu$.  So even
for the lowest Landau level, the peak is at $\omega \neq 0$, as we saw
in ref.\cite{Albash:2009wz}.  This lowest Landau level has
a boundary condition for the probe fermion at the horizon that is
constructed from the zeroth Hermite function. In this paper we will
study more general boundary conditions corresponding to the $n$th
Hermite function, and we find a family of quasiparticle peaks
signalling the presence of a Fermi surface at each such~$n$. Each
value of~$n$, for fixed chemical potential $\mu$, is a distinct Landau
level for the probe fermion.

There are two distinct classes of solution that we present and study
here. One class is separable (section~\ref{sec:separable}) and the other can be understood as an infinite sum of the separable solutions (section~\ref{sec:non-separable}). The first,
adopted in the work of refs.\cite{Basu:2009qz,Denef:2009yy}, does {\it
  not} have a smooth limit to the ${\cal H}=0$ case (for general
momentum $k_x$). As a result, we find it less compelling as a basis
for the generalization of the results and methods of
ref.\cite{Liu:2009dm} for the purposes of identifying Fermi surfaces by
searching for quasiparticle peaks, since the method requires that the
location (in ($\omega,k$)) of the peaks in the spectral function
$G_R(\omega,k)$ (generically appearing at non--zero $k_x$) be read off
from the system, and not fixed by hand. We stress that there is
nothing wrong with studying the separable class of solutions, for the
appropriate kind of physics of interest. Rather, we are noting that it
does not connect to the ${\cal H}=0$ physics in a way that allows an
analysis of the Fermi surface physics in the spirit of the prototype
case of ref.\cite{Liu:2009dm}. Following from this, we note in
section~\ref{sec:separable} that the choices made in
ref.\cite{Basu:2009qz}, when taken to their logical conclusion
(performing the required Fourier transform to construct the spectral
function), lead to no quasiparticle peaks at all.  By contrast, the infinite--sum class, which we began the study of in
ref.\cite{Albash:2009wz}, has a smooth limit to the ${\cal H}=0$ case,
as we discuss in section~\ref{sec:non-separable}, and we carry out the
search for quasiparticle peaks and display our results in
section~\ref{sec:numerical_results}.

%
\section{Free Fermions in a Magnetic Field}
\label{sec:freefermion}
To set the stage and for later comparison with the more complicated
case, we begin by studying a free fermion with mass $m$ charged under
a global $U(1)$.  The Dirac action is:
\begin{equation}
S_D = i \int d^3 x  \left( \bar{\Psi} \gamma^{\mu} \mathcal{D}_\mu \Psi - m \bar{\Psi} \Psi \right) \ ,
\end{equation}
where $\mathcal{D}_\mu = \partial_\mu - i q A_\mu$ is the covariant
derivative.  The classical equations of motion are given by:
\begin{equation}
\left( \gamma^{\mu} \mathcal{D}_\mu - m \right) \Psi = 0 \ .
\end{equation}
Let us first consider the case of zero magnetic field, taking only
$A_t = - \mu$ to be non--zero.  If we choose:
\begin{equation}
\gamma^0  = -i \sigma_3  \ , \quad \gamma^1 = \sigma_2 \ , \quad \gamma^2 = -\sigma_1 \ ,
\end{equation}
and take an ansatz of the form:
\begin{equation}
\Psi = e^{- i \omega t + i k_x x} \left( \begin{array}{c}  \phi_+ (\omega, k_x) \\ \phi_- (\omega, k_x) \end{array} \right) \ ,
\end{equation}
then the equations of motion reduce to (for $m = 0$):
\begin{equation}
\left( \omega - q \mu \right) \phi_\pm -  k_x \phi_\mp= 0 \ .
\end{equation}
The solutions are then given by:
\begin{eqnarray}
k_x = \omega - q \mu & \mathrm{with} & \phi_+ = \phi_- \ , \\
-k_x = \omega - q \mu & \mathrm{with} & \phi_+ = - \phi_- \ .
\end{eqnarray}
Note in particular that at $k_x = 0$, the non--trivial solution is
simply $ \omega - q \mu = 0$ with no restriction on $\phi_\pm(\omega,
k_x)$.

Let us now consider turning on a magnetic field.  Taking $A_x = - y
H$, and following a similar procedure as above (with $\phi_\pm$ now
depending on $y$ as well), we find as equations of motion:
\begin{equation}
\partial_y \phi_\pm  + m \phi_\mp \pm  \left( k_x + H q y \right) \phi_\mp = \left( \omega - q \mu \right) \phi_\mp  \ .
\end{equation}
Let us assume at present that $H q > 0$ and $m=0$ and define a new coordinate $\eta$ as:
\begin{equation} \label{eqt:eta}
\eta = \sqrt{H q} \left( y + \frac{k_x}{H q} \right) \ , 
\end{equation}
in terms of which the equation of motion  simplifies to:
\begin{equation} \label{eqt:eom}
\sqrt{H q} \left(\partial_\eta \phi_\pm \pm   \eta \phi_\mp \right) =  \pm  \left( \omega - q \mu \right) \phi_\mp  \ .
\end{equation}
If we consider solutions of the form:
\begin{equation}
\phi_+  = c(\omega,k_x) I_n(\eta) \ , \quad \phi_-  = c(\omega,k_x) I_{n-1} (\eta) \ , 
\end{equation}
where $I_n$ are the standard Hermite functions defined in terms of the
Hermite polynomials $H_n$ as :
\begin{equation}
\label{eq:hermite}
I_n(\eta) \equiv N_n H_n= \frac{1}{\sqrt{2^{n} n! \sqrt{\pi}}} e^{-\eta^2/2} H_{n}(\eta) \ ,
\end{equation}
 we have the requirement that:
\begin{equation}
\sqrt{ 2 H q n} = \omega - q \mu  \ .
\end{equation}
Another solution is given by:
\begin{equation}
\phi_+  = c(\omega,k_x) I_n(\eta) \ , \quad \phi_-  = -c(\omega,k_x) I_{n-1} (\eta) \ , 
\end{equation}
The solutions for the fields are the same as before, but now we have the requirement:
\begin{equation}
-\sqrt{ 2 H q n} = \omega - q \mu  \ .
\end{equation}
%
Let us now consider the case of $H q < 0$.  The analysis is mostly the
same, except that we now define $\eta$ as:
\begin{equation}
\eta = -\sqrt{-H q} \left( y + \frac{k_x}{H q} \right)
\end{equation}
This modifies the equation of motion to:
\begin{equation}
\sqrt{-H q} \left(\partial_\eta \phi_\pm \mp  \eta \phi_\pm \right) =  \mp \left( \omega - q \mu \right) \phi_\mp \ .
\end{equation}
The solutions to this equation are given by:
\begin{equation} \label{eqt:sol3}
\left\{ \begin{array}{c} \phi_+(\omega,k_x,y) = c(\omega,k_x) I_{n-1}(\eta)  \ , \quad \phi_-(\omega,k_x,y) = c(\omega,k_x)  I_n(\eta)  \ , \\ \\
\sqrt{-2 H q n} = \omega - \mu q  \ ,
\end{array} \right.
\end{equation}
and
\begin{equation} \label{eqt:sol4}
\left\{ \begin{array}{c} \phi_+(\omega,k_x,y) = c(\omega,k_x) I_{n-1}(\eta)  \ , \quad \phi_-(\omega,k_x,y) = -c(\omega,k_x)  I_n(\eta)  \ , \\ \\
-\sqrt{-2 H q n} = \omega - \mu q  \ .
\end{array} \right.
\end{equation}
In particular, we note that flipping the sign on the magnetic field
interchanges the $\eta$ dependence for $\phi_\pm$.

Finally, let us study the zero magnetic field limit, which is a  delicate issue.  We first point out that the coordinate change presented in equation \reef{eqt:eta} is singular in the limit of zero magnetic field, so it is not a particularly good basis for studying this limit.  However, one might think that in order for the restriction on $\omega$ to
match, we have to choose:
\begin{equation} \label{eqt:limit1}
\lim_{H \to 0} \left(\pm \sqrt{ 2 |H q| n}\right) = \pm k_x = \mathrm{fixed} \ .
\end{equation}
However, in this limit, the functions $I_n$ behave as:
\begin{equation}
I_n(\eta) \sim \pi^{- \frac{1}{2}} 2^{n- \frac{1}{4}} n^{-\frac{1}{4}}e^{\mp k_x y - \frac{n}{2}} \ ,
\end{equation}
so both $\phi_+$ and $\phi_-$ vanish in the limit of large $n$.  This
does not match the zero magnetic field solution we found earlier.  We can perhaps compensate for this vanishing by including an $n$--dependent overall normalization of our fields $\phi_\pm$, but we emphasize that this issue is not present when $k_x = 0$. There, no special limit
needs to be taken on $n$ and the Hermite functions go to a constant in
the limit of $H \to 0$.  This sort of property will occur in our full
probe fermion problem later, for one class of solutions (separable)
that we will discuss. There will be another class of solutions
(infinite--sum) that will not suffer from this limitation, having a
smooth $H=0$ limit for arbitrary $k_x$. This latter class is naturally
amenable to being used to search for Fermi surfaces by looking for
quasiparticle peaks in the spectral function, since these peaks will
naturally occur at non--zero $k_x$ which we wish to read off as
output, and not fix {\it a priori}.

%
%
\section{Probing the Black Hole}
Our probe is a Dirac fermion in an asymptotically AdS$_4$ dyonic black
hole background.  It is charged under the background's $U(1)$, and the
metric and fields are given by:
\begin{eqnarray} \label{eqt:dyonic_metric}
ds^2 &=& \frac{L^2 \alpha^2}{z^2} \left( - f \left(z \right) dt^2 + dx^2 + d y^2 \right) +\frac{L^2}{z^2} \frac{d z^2}{f \left(z \right)} \ ,  \\
F &=& 2 H \alpha^2  dx \wedge d y + 2 Q \alpha dz \wedge dt\ , \nonumber \\
f \left(z \right) &=& 1 + \left( H^2 + Q^2 \right) z^4 - \left(1 + H^2 + Q^2 \right) z^3 = \left(1-z \right)\left( z^2 +z+ 1 - \left(H^2 +Q^2\right) z^3 \right) \ . \nonumber
\end{eqnarray}
The parameter $\alpha$ has dimensions of inverse length. The negative
cosmological constant sets the length scale $L$, $\Lambda=-3/L^2$,  and
the Einstein--Maxwell action is:
\begin{equation}
\label{eq:emads}
S_{\rm bulk}=\frac{1}{2\kappa_4^2}\int \! d^4x \sqrt{-G}\left\{ R+\frac{6}{L^2}-\frac{L^2}{4}F^2\right\}\ ,
\end{equation}
where we use signature $(-+++)$, and $\kappa^2_4=8\pi G_N$. The mass
per unit volume and temperature of the background are given by:
\begin{equation}
  \varepsilon=\frac{\alpha^3 L^2}{\kappa^2_4}[1+Q^2+H^2]\ ,\quad T=\frac{\alpha}{4\pi}[3-(Q^2+H^2)]\ .
\end{equation}
The coordinates we have chosen above in
equation~\reef{eqt:dyonic_metric} are such that the boundary of
AdS$_4$ (where the dual $(2+1)$--dimensional theory is defined in the
ultraviolet) is at $z=0$ while the horizon of the black hole is at
$z=1$.

We choose a gauge such that:
\begin{equation}
A_t=2Q\alpha(z-1)\ ,\quad \mathrm{and} \quad  A_x=-2H\alpha^2 y\ ,
\end{equation}
which sets a chemical potential $\mu=-Q\alpha$ and a magnetic field
${\cal H}=-2H\alpha^2$.  We choose to work at zero temperature, which
restricts $Q$ and $H$ to satisfy:
\begin{equation}
H^2 + Q^2 = 3 \ .
\end{equation}
As discussed in our first paper\cite{Albash:2009wz}, it is
important to realize that this equation does {\it not} place a
restriction on the magnetic field for a given value of $Q$. The
important quantity to hold fixed while varying the magnetic field is
the chemical potential $\mu$, and as $Q$ decreases for increasing $H$,
$\alpha$ increases to compensate, resulting in the physical magnetic
field ${\cal H}$ being able to run its full natural range from zero to
infinity.

Now $z$ is dimensionless in our equations above while all the other
coordinates are dimensionful. For convenience, in what follows we will
rescale our fields and coordinates to dimensionless quantities:
\begin{equation}
  t\to t/\alpha\ ,\quad x\to x/\alpha\ ,\quad y\to y/\alpha\ ,\quad A_t\to \alpha A_t\ ,\quad\mathrm{and}\,\, A_x\to \alpha^2 A_x\ .
\end{equation}
The Dirac action for our probe fermion is:
\begin{equation}
\label{eq:dirac}
S_D=i\!\int \! d^4 x\sqrt{-G}\left({\bar\Psi}\Gamma^M{\cal D}_M\Psi-m{\bar \Psi}\Psi\right) \ ,
\end{equation}
with the covariant derivative ${\cal D}_M$ given by:
\begin{equation}
{\cal D}_M=\partial_M+\frac14\omega_{abM}\Gamma^{ab}-iqA_M\ ,
\end{equation}
where we have used vielbeins $e^M_a$ to exchange  curved spacetime indices $\{M, N\}$ for
tangent space indices $\{a,b\}$, and 
\begin{equation}
  \Gamma^{ab}\equiv\frac12[\Gamma^a,\Gamma^b]\ ,\quad \omega_{abM}\equiv e^{N}_a\partial_M e_{bN}-e_{a\, N}e^O_b\Gamma^N_{OM}\ .
\end{equation}
The fermion couples to an operator of dimension $\Delta=m+d/2$ in the
$(2+1)$--dimensional theory. We will later choose the case $m=0$ for
much of the treatment we will do in this paper.  We choose for our
Gamma matrices:
\begin{eqnarray}
\Gamma^t &=& \left( \begin{array}{cc}
- i \sigma_2 & 0 \\
0 & - i \sigma_2
\end{array} \right) \ , \quad
\Gamma^x = \left( \begin{array}{cc}
-  \sigma_3 & 0 \\
0 & \sigma_3
\end{array} \right) \ , \nonumber \\
\Gamma^y&=& \left( \begin{array}{cc}
0 & - i \sigma_3 \\
i \sigma_3 & 0
\end{array} \right) \ , \quad
\Gamma^z= \left( \begin{array}{cc}
-  \sigma_1 & 0 \\
0 & -\sigma_1
\end{array} \right)\ ,
\end{eqnarray}
where the $\sigma_i$ are the standard Pauli matrices.

%
%
We write the Dirac spinor in terms of two 2--component spinors:
\begin{equation}
\Psi^T =  z^{3/2} f(z)^{-1/4} e^{- i \omega t + i k_x x}  \left( \phi_1 \ , \phi_2 \right) \ ,
\end{equation}
and the equation of motion reduces to:
\begin{equation}
\sqrt{\frac{g_{xx}}{g_{zz}}} \left(\partial_z \phi_{^1_2} + \frac{m}{z} \sigma_1 \phi_{^1_2}\right) + i u \sigma_3 \phi_{^1_2} \pm \sigma_2 \left( \partial_y \phi_{^2_1} + \left( k_x + 2 H q y \right) \phi_{^1_2} \right) = 0\ ,
\end{equation}
with:
\begin{equation}
u = \sqrt{ \frac{g_{xx}}{-g_{tt}}} \left( \omega + 2 q Q (z-1) \right) \ , \ \sqrt{ \frac{g_{xx}}{g_{zz}}} = \sqrt{f} \ ,  \sqrt{ \frac{g_{xx}}{-g_{tt}}} = \frac{1}{\sqrt{f}} \ .
\end{equation}
Near the AdS boundary, the solutions to this equation asymptote to:
\begin{equation}
\lim_{z \to 0} \phi_{^1_2} (y,z)= a_{^1_2}(y) \left( \begin{array}{c} 1 \\ 1 \end{array} \right) z^{-m} + b_{^1_2}(y) \left( \begin{array}{r} -1 \\ 1 \end{array} \right) z^m\ .
\end{equation}
We can define 4--component spinors $\phi_\pm$ that have eigenvalues $\pm 1$ under $\Gamma^z$:
\begin{equation}
\phi_\pm (y,z)= \frac{1}{2} \left( 1 \pm \Gamma^z \right) \Psi(y,z) = \frac{1}{2} \left( \begin{array}{c} \left(1 \mp \sigma_1 \right) \phi_1(y,z) \\ \left( 1 \mp \sigma_1 \right) \phi_2(y,z) \end{array} \right)\ .
\end{equation}
In the limit of $z \to 0$, $\phi_\pm$ asymptote to:
\begin{equation}
\phi_+(y,z) = z^m \left( \begin{array} {r} -b_1 \\ b_1 \\ - b_2 \\ b_2 \end{array} \right) \ , \quad \phi_-(y,z) = z^{-m} \left( \begin{array} {r} a_1 \\ a_1 \\ a_2 \\ a_2 \end{array} \right) \ .
\end{equation}
Using the prescription of ref.~\cite{Iqbal:2009fd} to calculate the
retarded Green function (the spectral function of interest), we write:
\begin{equation}
\left( \begin{array} {r} -\tilde{b}_1 \\ \tilde{b}_1 \\ - \tilde{b}_2 \\ \tilde{b}_2 \end{array} \right) ={\cal S}\left( \begin{array} {r} \tilde{a}_1 \\ \tilde{a}_1 \\ \tilde{a}_2 \\ \tilde{a}_2 \end{array} \right)= \left( \begin{array}{cccc}
 0 &  - \frac{\tilde{b}_1}{\tilde{a}_1} & 0 & 0 \\
 \frac{\tilde{b}_1}{\tilde{a}_1} & 0 & 0 & 0 \\
 0 & 0 & 0 & - \frac{\tilde{b}_2}{\tilde{a}_2} \\
 0 & 0 &  \frac{\tilde{b}_2}{\tilde{a}_2} & 0 \end{array} \right) \left( \begin{array} {r} \tilde{a}_1 \\ \tilde{a}_1 \\ \tilde{a}_2 \\ \tilde{a}_2 \end{array} \right) \ ,
\end{equation}
where $\tilde{a}$ and $\tilde{b}$ are the Fourier transform in the
$y$--direction of $a$ and $b$ respectively and hence find for the
retarded Green function:
\begin{equation}
G_R = - i \mathcal{S} \Gamma^t = i \left( \begin{array}{cccc}
  \frac{\tilde{b}_1}{\tilde{a}_1} &  0 & 0 & 0 \\
0 &  \frac{\tilde{b}_1}{\tilde{a}_1} & 0 & 0 \\
 0 & 0 & \frac{\tilde{b}_2}{\tilde{a}_2} & 0 \\
 0 & 0 & 0 &  \frac{\tilde{b}_2}{\tilde{a}_2} \end{array} \right)\ .
 \end{equation}
 The dual (boundary) field theory should have half the components that
 the bulk theory has, and this is reflected in the Green function by
 having only two independent terms.  Finally, if we write:
\begin{equation}
  \phi_1 = \left( \begin{array}{c} A_1 \\ B_1 \end{array} \right) \ ,\quad \mathrm{and}\quad   \phi_2 = \left( \begin{array}{c} A_2 \\ B_2 \end{array} \right)\ ,
\end{equation}
then the functions $(a_{^1_2}, b_{^1_2})$ can be written:
\begin{equation}
a_{^1_2}(y) = \lim_{\epsilon \to 0} \frac{\epsilon^m}{2} \left(B_{^1_2}(y,\epsilon) + A_{^1_2}(y,\epsilon) \right) \ , \quad  b_{^1_2}(y) = \lim_{\epsilon \to 0} \frac{\epsilon^{-m}}{2} \left(B_{^1_2}(y,\epsilon) - A_{^1_2}(y,\epsilon) \right)\ ,
\end{equation}
(with analogous expressions for the Fourier transforms $({\tilde a},
{\tilde b})$, with $({\tilde A},{\tilde B})$ instead of $(A,B)$) such
that the relevant Green function quantities are simply:
\begin{equation} \label{eqt:Green}
G_R^{(1)}(\omega, k_x,k_y)  =  \lim_{\epsilon \to 0} i \epsilon^{-2 m} \frac{\tilde{B}_1 - \tilde{A}_1}{ \tilde{B}_1 +  \tilde{A}_1} \ , \quad G_R^{(2)} (\omega, k_x,k_y)= \lim_{\epsilon \to 0}  i \epsilon^{-2 m} \frac{ \tilde{B}_2 -  \tilde{A}_2}{ \tilde{B}_2 +  \tilde{A}_2}\ .
\end{equation}
To solve for the fields $A_{^1_2}, B_{^1_2}$, we define:
\begin{equation}
A_{^1_2} =  e^{i \omega/(6 (1-z))}   (1-z)^{i (6 q Q - 4 \omega)/18} A_\pm \  , \quad B_{^1_2} = e^{i \omega/(6 (1-z))}   (1-z)^{i (6 q Q - 4 \omega)/18} B_\pm \ .
\end{equation}
The equations of motion for the fields $(A_\pm, B_\pm)$ when  $m = 0$ is given by:
\begin{eqnarray}
\sqrt{\frac{g_{xx}}{g_{zz}}} \left(\partial_z  + \frac{ i \omega}{6 (1-z)^2}   + i \frac{- 6 q Q + 4 \omega}{18 (1 - z)}  \right) A_+&=& - i u A_+ + i \left( \partial_y B_- + (2H q y + k_x) B_+ \right)  , \nonumber \\
\sqrt{\frac{g_{xx}}{g_{zz}}} \left( \partial_z  + \frac{ i \omega}{6 (1-z)^2}   + i \frac{- 6 q Q + 4 \omega}{18 (1 - z)}  \right) A_-&=& - i u A_-  - i \left( \partial_y B_+ + (2H q y + k_x) B_- \right)  , \nonumber \\
\sqrt{\frac{g_{xx}}{g_{zz}}} \left( \partial_z  + \frac{ i \omega}{6 (1-z)^2}   + i \frac{- 6 q Q + 4 \omega}{18 (1 - z)}  \right) B_+ &=&  + i u B_+ -i  \left( \partial_y A_- + (2H q y + k_x) A_+ \right)  , \nonumber \\
\sqrt{\frac{g_{xx}}{g_{zz}}} \left( \partial_z  + \frac{ i \omega}{6 (1-z)^2}   + i \frac{- 6 q Q + 4 \omega}{18 (1 - z)}  \right) B_- &=&  + i u B_- + i \left( \partial_y A_+ + (2H q y + k_x) A_- \right)  \ .\nonumber\\
\end{eqnarray}
These are the equations of motion we presented in
ref.~\cite{Albash:2009wz}.  Expanding the equations of motion at the
event horizon, we find (for $\omega \neq 0$) the following conditions:
\begin{eqnarray}
A_{\pm} (y,1) &=& 0 \ , \\
\partial_z A_\pm (y,1) &=& \mp \frac{\sqrt{6}}{2 \omega} \left( k_x B_\pm(y,1) +  \partial_y B_\mp(y,1) + 2 H q y B_\pm(y,1) \right) \ ,  \\
\partial_z B_\pm (y,1) &=& \frac{-i}{108} \left( \mp 18 \sqrt{6} \left( k_x \partial_z A_\pm(y,1) + \partial_z \left( \partial_y A_\mp(y,1) + 2 H q y A_\pm(y,1) \right) \right) \right. \nonumber \\
&& \left. +  \left(48 q Q - 23 \omega \right) B_\pm(y,1) \right) \ . 
\end{eqnarray}
(There are analogous equations for $\omega=0$, but we do not list them
here since we will not find quasiparticle peaks at $\omega=0$.)  There
is some freedom in choosing the boundary conditions for the fields,
leading to very different physics.  We present two particular choices
in the next two sections.
%
\subsection{Separable Solutions}
\label{sec:separable}
%
We restrict ourselves at present to $q H > 0$.  Changing coordinates to:
\begin{equation}
\eta = \sqrt{2 H q} \left(y + \frac{k_x}{2 H q} \right) \ ,
\end{equation}
we can write the $y$--dependent parts of our equations above as:
\begin{eqnarray}
\partial_y B_\pm + (2H q y + k_x) B_\mp &=& \sqrt{2 H q} \left( \partial_\eta B_\pm +\eta B_\mp \right) \ , \\
\partial_y A_\pm + (2H q y + k_x) A_\mp &=& \sqrt{2 H q} \left( \partial_\eta A_\pm + \eta A_\mp \right) \ .
\end{eqnarray}
There are two possible ans\"atze that we can make:
\begin{equation}
\mathrm{ansatz \ 1}:  \ A_- = - A_+ = - Y_A(y) Z_A(z) \ , \quad B_- = B_+ = Y_B(y) Z_B(z) \ ,
\end{equation}
or
\begin{equation}
\mathrm{ansatz \ 2}: \ A_- =  A_+ =  Y_A(y) Z_A(z) \ , \quad B_- = - B_+ = - Y_B(y) Z_B(z) \ .
\end{equation}
We emphasize that the ansatz with $A_- = A_+$ and $B_- = B_+$ is not consistent with the first order equations.
Let us start with the first ansatz.  If we write:
\begin{eqnarray}
\partial_\eta Y_B +\eta Y_B &=& \sqrt{2 n} Y_A \ , \\
\partial_\eta Y_A - \eta Y_A &=& - \sqrt{2 n} Y_B \ ,
\end{eqnarray}
then the solution takes the form:
\begin{equation}
Y_A = I_{n-1}(\eta) \ , \quad Y_B =  I_n(\eta) \ ,
\end{equation}
where $I_n$ is the Hermite function defined in
section~\ref{sec:freefermion}.  Note that $I_{-1} = 0$.  With this
ansatz\footnote{These are the separable solutions discussed in
  refs.~\cite{Basu:2009qz, Denef:2009yy}, and they are somewhat
  analogous to the standard separable free fermion case reviewed in
  section~\ref{sec:freefermion}. We considered them in our work in
  ref.\cite{Albash:2009wz} for $n=0$, but focused on the infinite--sum
  solutions for the rest of that paper. The lack of a smooth ${H=0}$
  limit for $k_x\neq0$, combined with the desire to seek quasiparticle
  peaks at non--zero $k_x$, led us away from these separable cases. We
  will discuss this more below.}, the equations of motion in the bulk
become:
\begin{eqnarray}
\label{eq:become}
\sqrt{\frac{g_{xx}}{g_{zz}}} \left(\partial_z  + \frac{ i \omega}{6 (1-z)^2}   + i \frac{- 6 q Q + 4 \omega}{18 (1 - z)}  \right) Z_A&=& - i u Z_A+2  i \sqrt{H q n} Z_B \ , \nonumber\\
\sqrt{\frac{g_{xx}}{g_{zz}}} \left( \partial_z  + \frac{ i \omega}{6 (1-z)^2}   + i \frac{- 6 q Q + 4 \omega}{18 (1 - z)}  \right) Z_B &=&  + i u Z_B - 2 i   \sqrt{H q n} Z_A \ .
\end{eqnarray}
If we consider the second ansatz, we can write:
\begin{eqnarray}
\partial_\eta Y_B -\eta Y_B &=& -\sqrt{2 n} Y_A \ , \\
\partial_\eta Y_A + \eta Y_A &=&  \sqrt{2 n} Y_B \ ,
\end{eqnarray}
which has solutions given by:
\begin{equation}
Y_A (y)= I_n(\eta) \ , \quad Y_B(y) = I_{n-1} (\eta) \ ,
\end{equation}
and the equations of motion reduce to equations~\reef{eq:become}, the
same set of equations of motion that resulted from the first ansatz.
So in this case, the behavior in the $z$ direction is independent of
our choice. We will return to the issue of this freedom later.
%

We proceed with calculating the Green function given by equation
\reef{eqt:Green}, where all the fields have been Fourier transformed
in both $(x,y)$ coordinates.  Note that in the analysis of
ref.~\cite{Basu:2009qz}, no Fourier transform was performed for the
$y$ coordinate.  The Fourier transform of the Hermite functions is
given by:
\begin{equation}
\int_{-\infty}^{\infty} dy e^{-i k_y y} I_n (\eta) = e^{ i \frac{k_x}{\sqrt{2 H q}}}\sqrt{\frac{2 \pi}{2 H q}} N_n (-i)^n e^{-\left(\frac{k_y}{\sqrt{2 H q}} \right)^2 / 2} H_n \left(\frac{k_y}{\sqrt{2 H q}} \right) \ ,
\end{equation}
where the normalization $N_n$ was given in equation~\reef{eq:hermite}.

Let us compare to the results and methods of ref.~\cite{Liu:2009dm},
where $H=0$. There, the choice $k_y=0$ is made, and we can examine the
physics of the solution we have here at $H\neq0$ for $k_y = 0$ to see
how it connects.  If we restrict to $k_y=0$, we find that for $n$ odd,
the Fourier transform of $I_n(\eta)$ vanishes.  Therefore, when $n$ is
even, only $\tilde{B}_\pm(k_x, k_y =0)$ survive when using the first
ansatz, whereas only $\tilde{A}_\pm(k_x, k_y=0)$ survive when using the
second ansatz.  When instead $n$ is odd, only $\tilde{A}_\pm(k_x,
k_y=0)$ survive for the first ansatz and only $\tilde{B}_\pm(k_x,
k_y=0)$ survive when using the second ansatz.  Therefore, if we wish our
retarded Green function to be positive, we must use the first ansatz
to describe even modes and the second ansatz to describe odd modes.
This gives:
\begin{equation} \label{eqt:trivialGreen}
G_R(\omega, k_x , k_y = 0)^{(1),(2)} =  i \ .
\end{equation}
Let us now consider the $H \to 0$ limit of our solutions.  As shown
earlier for the free fermion in 2+1 dimensions, the Hermite functions
in terms of $\eta$ do not have nice behavior as $H \to 0$.
Furthermore, the Green function presented above in the $H \to 0$ limit
cannot match the results presented in ref.~\cite{Liu:2009dm}, since
their results at $H = 0$ are a constant only at $k_x = 0$.  Therefore,
these results suggest that this choice of solutions (used, crucially,
in ref.\cite{Basu:2009qz} to discuss Fermi surfaces) is not connected
to the pole found in ref.~\cite{Liu:2009dm} at $H=0$, but they form a
separate set of solutions present at non--zero magnetic field.

We complete our discussion of this set of solutions by considering the
case of $q H < 0$.  The coordinate change is now given by:
\begin{equation}
\eta = - \sqrt{ - 2 H q} \left( y + \frac{k_x}{2 H q} \right) \ . 
\end{equation}
Under the change of variables, the $y$--dependent part becomes:
\begin{eqnarray}
\partial_y B_\pm + (2H q y + k_x) B_\mp &=&- \sqrt{-2 H q} \left( \partial_\eta B_\pm -\eta B_\mp \right) \ , \nonumber \\
\partial_y A_\pm + (2H q y + k_x) A_\mp &=&- \sqrt{-2 H q} \left( \partial_\eta A_\pm - \eta A_\mp \right) \ .
\end{eqnarray}
If we now consider our two ans\"atz from before, we find that ansatz 1 results in:
\begin{equation}
B_+ (y,z)= B_-(y,z) = Z_B(z) I_{n-1}(\eta) \ , \quad A_+(y,z) = - A_-(y,z) = Z_A(z) I_{n}(\eta) \ ,
\end{equation}
and ansatz 2 gives:
\begin{equation}
B_+(y,z) = - B_-(y,z) = Z_B(z) I_{n}(\eta) \ , \quad A_+ (y,z)=  A_-(y,z) = Z_A(z) I_{n-1}(\eta) \ .
\end{equation}
Both these ans\"atze give exactly the same equation of motion as
before with the appropriate modification under the square root:
\begin{eqnarray}
\sqrt{\frac{g_{xx}}{g_{zz}}} \left(\partial_z  + \frac{ i \omega}{6 (1-z)^2}   + i \frac{- 6 q Q + 4 \omega}{18 (1 - z)}  \right) Z_A&=& - i u Z_A +2  i \sqrt{-H q n} Z_B \ , \nonumber \\
\sqrt{\frac{g_{xx}}{g_{zz}}} \left( \partial_z  + \frac{ i \omega}{6 (1-z)^2}   + i \frac{- 6 q Q + 4 \omega}{18 (1 - z)}  \right) Z_B &=&  + i u Z_B - 2 i   \sqrt{-H q n} Z_A \ .
\end{eqnarray}
In turn this suggests that when $H q < 0$, the even modes are
associated with ansatz 2 and the odd modes are associated with ansatz
1.  This is the opposite of what occurred when $ H q > 0$.  

\noindent Let us summarize our results so far:
\begin{equation}
q H > 0 \left\{ \begin{array}{c} \mathrm{ansatz} \  1 (n\,\,{\rm even}) \left(\begin{array}{c} Z_A I_{n-1} \\ Z_B I_n \\ - Z_A I_{n-1} \\ Z_B I_n \end{array} \right) \\ \\
 \mathrm{ansatz} \  2 (n\,\,{\rm odd}) \left(\begin{array}{c} Z_A I_{n} \\ Z_B I_{n-1} \\ Z_A I_{n} \\ -Z_B I_{n-1} \end{array} \right) 
\end{array} \right. \ , \
 q H < 0 \left\{ \begin{array}{c} \mathrm{ansatz} \  1 (n\,\,{\rm odd}) \left(\begin{array}{c} Z_A I_{n} \\ Z_B I_{n-1} \\ - Z_A I_{n} \\ Z_B I_{n-1} \end{array} \right) \\ \\
 \mathrm{ansatz} \  2 (n\,\,{\rm even}) \left(\begin{array}{c} Z_A I_{n-1} \\ Z_B I_{n} \\ Z_A I_{n-1} \\ -Z_B I_{n} \end{array} \right) 
\end{array} \right. \ .
\end{equation}
Note the close similarity between these results and the results found
for the free fermion in $2+1$ dimensions.  We find here that for a
given ansatz, the single Dirac spinor in the $(3+1)$--dimensional
background contains both solutions that were found for the free
fermion.  Under the flip of the magnetic field for a given ansatz, we
find the same flipping of the $\eta$ dependence between each
2--component spinors.  In addition, the flipping of the magnetic
field interchanges the roles of the two ans\"atze for describing the
modes, which suggests that they may describe aligned {\it vs.}
anti--aligned couplings.  These results indicate that the even modes
and odd modes should be understood as describing two distinct ladders.
%
\subsection{Infinite--Sum Solutions}
\label{sec:non-separable}
%
The separable solutions presented so far have two unfortunate features.  Because of the solutions' dependence on the coordinate $\eta$, the solutions have singular behavior in the zero magnetic field limit.  To connect these solutions to the zero magnetic field solution, one has to take a limit similar to that of equation \reef{eqt:limit1}, where the Landau level label $n$ goes to infinity:
\begin{equation} \label{eqt:limit2}
\lim_{H \to 0, n \to \infty} \sqrt{4 H q n} = k_x
\end{equation}
Independent of this fact, considering the separable solutions alone leads to a trivial Green function, as presented in equation \reef{eqt:trivialGreen}.

We wish to proceed by studying a class of solutions that avoids these two features.  In order to avoid singularities associated with the coordinate $\eta$ , we wish to consider solutions that depend on the coordinate $y$ as opposed to $\eta$.  Let us take the case of $H q > 0$.  Motivated by the separable solutions discussed earlier, focusing on the behaviour at the horizon ($z=1$), we consider $B_\pm(y,1) = B(y,1)$ for even $n$ and $B_+ =
-B_- = B(y,1)$ for odd $n$.  For $n$ even we take:
\begin{equation}
\partial_y B(y,1) + 2 H q y B(y,1) = \sqrt{4 H q n} B_0(1) I_{n-1}(\sqrt{2 H q} y) \ ,
\end{equation}
such that we have $B(y,1) = B_0(1) I_n(\sqrt{2 H q} y)$.  We can write this ansatz in terms of our separable solutions from the previous section,
\begin{equation} \label{eqt:series}
B_+(y,z) = \sum_{n = 0}^{\infty} b_+^{(n)} Z_B^{(n)}(z) I_n(\eta) \ ,
\end{equation}
and similarly for the other fields.  The coefficients $b_+^{(n)}$ can be determined from the field behavior at the event horizon.  For example, for $B(y,1) = I_0( \sqrt{2 H q} y)$, the coefficients $b_+^{(n)}$ are given by:
\begin{equation}
b_+^{(n)} = \frac{k_x^n}{2^n \sqrt{\left( H q \right)^n n!}}e^{- \frac{k_x^2}{8 H q}} \ .
\end{equation}
In the limit of $H \to 0$, the infinite sum remains finite  \emph{without} requiring us to take a limit of the form of equation \reef{eqt:limit2}.  We do not need to take $n \to \infty$ since the infinite sum effectively does this by defining a new Landau level labeling in terms of an infinite sum of the Landau labels of the separable solutions (\emph{i.e.} the labeling of $I_n( \sqrt{2 H q} y)$ instead of $I_n (\eta)$).  Because of the nature of the infinite sum, it is important to first perform the sum and then take the limit of $H \to 0$ to ensure the correct result.  We emphasize that the infinite sum gives us the freedom to vary the new Landau level label as well as $k_x$ independently while sending $H \to 0$, a freedom we did not have with the separable solutions presented earlier.  This is an important difference from the separable solutions we considered in the previous section.

By substituting the series form of the fields from equation \reef{eqt:series} into the equations of motion, the partial differential equation becomes an infinite series of coupled ordinary differential equations for $Z_{A,B}^{(n)}$, where the coupled equations are identical to those of equations \reef{eq:become}.  We have traded the partial differential equation for an infinite number of coupled ordinary differential equations.  We choose to proceed by solving the partial differential equation using an ansatz that is not explicitly separable since this is more tractable numerically.

With this ansatz, we find that:
\begin{equation} \label{eqt:bc1}
\partial_z A_{\pm} (y,1) = \mp \frac{\sqrt{6}}{2 \omega} \left( k_x I_n (\sqrt{2 H q} y) + \sqrt{4 H q n} I_{n-1} (\sqrt{2 H q} y) \right) B_0(1) \ .
\end{equation}
We notice that for $k_x = 0$, we find that $\partial_z A_\pm(y,1) \propto I_{n-1}(\sqrt{2 H q} y)$, which agrees with the separable solution from the previous section.  Finally, we have:
\begin{eqnarray} 
\partial_z B_{\pm} &=& \frac{-i B_0(1)}{108} \left(  \frac{54}{\omega} \left( k_x^2 I_n + k_x \sqrt{ 4 H q n} I_{n-1} + k_x \sqrt{4 H q (n+1)} I_{n+1} + 4 H q n I_n\right) \right. \nonumber \\
&& \left.+ \left( 48  q Q -23  \omega  \right) I_n \right) \ , \label{eqt:bc2}
\end{eqnarray}
where we have used that:
\begin{equation}
\partial_y I_n ( \sqrt{ 2 q H} y) - 2 q H y I_n ( \sqrt{ 2 q H} y)  = - \sqrt{4 q H (n+1)} I_{n+1}  \ .
\end{equation}
Note again that for $k_x = 0$, we recover that $\partial_z B_\pm(y,1)
\propto I_n( \sqrt{2 H q} y)$ which is the single separable solution
presented earlier.  In addition, note that for $n=0$ we recover the
equations that we had in ref.~\cite{Albash:2009wz}.  For $n$ odd, we
take:
\begin{equation}
\partial_y B(y,1) - 2 H q y B(y,1) = - \sqrt{4 H q n} B_0(1) I_{n}(\sqrt{2 H q} y) \ ,
\end{equation}
such that we have $B(y,1) = B_0(1) I_{n-1}(\sqrt{2 H q} y)$.  With this ansatz, we find that:
\begin{eqnarray}
\partial_z A_{\pm} (y,1) &=& - \frac{\sqrt{6}}{2 \omega} \left( k_x I_{n-1} (\sqrt{2 H q} y) + \sqrt{4 H q n} I_{n} (\sqrt{2 H q} y) \right) B_0(1) \ , \nonumber\\
\partial_z B_{\pm} &=& \mp \frac{i B_0(1)}{108} \left(  \frac{54}{\omega} \left( k_x^2 I_{n-1} + k_x \sqrt{ 4 H q n} I_{n} + k_x \sqrt{4 H q (n-1)} I_{n-2} + 4 H q n I_{n-1}\right) \right. \nonumber \\
&& \left.+ \left( 48  q Q -23  \omega  \right) I_{n-1} \right) \ . \label{eqt:bctwo}
\end{eqnarray}
The advantage of this choice over the finitely separable choice from the
previous section is that in the $H \to 0$ limit, the equations of
motion and fields reduce to those of ref.\cite{Liu:2009dm} with no
restrictions placed on $n$.  In addition, since at $k_x = 0$, the
fields reduce to the simple separable case discussed earlier, we can
immediately write the Green function for this special case:
\begin{equation}
G_R(\omega, k_x = 0, k_y = 0)^{(1),(2)} = i \ ,  \qquad \forall H \ ,
\end{equation}
which connects nicely with the zero magnetic field ${\vec k}=0$ result
of ref.\cite{Liu:2009dm}.  Furthermore, for $k_x \neq 0, \ k_y = 0$, the Green function gives a non--trivial result, as opposed to the separable case which gave a trivial result (see equation \reef{eqt:trivialGreen}).

This infinite--sum class of solutions, which we studied in
ref.\cite{Albash:2009wz} at $n=0$, is a more natural generalization of
the results of ref.\cite{Liu:2009dm} and better adapted to the method
of searching for new quasiparticle peaks, which generically appear at
non--zero $k_x$.
%
\subsection{Comments about Complex $\omega$}
%
Before proceeding with our numerical results, we would like to make a
few comments about the variable $\omega$.  In the usual AdS/CFT
dictionary, it is important to recall that $\omega$ is generally
 complex:
\begin{equation}
\omega = \omega_\ast - i \Gamma
\end{equation}
The quantity $\Gamma$ is related to the inverse lifetime of the
quasinormal mode and $\omega_\ast$ is related to the energy of the
mode.  The imaginary part of $\omega$ must be negative in order for
the mode to be stable.  In addition, in order for the pole in the
Green function to be associated with a quasiparticle, the pole must
have non--zero residue.  In our analysis, we restrict ourselves to
$\Gamma = 0$, and therefore our search for poles is such that the
lifetime of the mode is always infinite.  Away from the poles, the
numerical results should not be taken too seriously since away from
the poles, one should consider a complex $\omega$ and not a purely
real~$\omega$.  This is relevant because in our numerics the poles
occur immediately after a region where the imaginary part of the
Green function is negative, which is not allowed if the theory is
unitary.  We believe that the proper treatment of the complex $\omega$
in these regions would resolve this.  In particular, the analysis in
ref.~\cite{Denef:2009yy} shows that the pole for $H=0$ bounces off the
real axis, and in the purely real analysis the imaginary part of the
Green function remains positive on both sides of the pole.
Therefore, our result suggests that the bounce does not occur at the
real axis but somewhere in the positive imaginary $\omega$ regime.
This suggests that for $\mathrm{Re}(\omega)$ less than our pole, the
mode is unstable.
In addition, we have not calculated the residue of our poles as this
is impossible without having the full complex $\omega$ description of
the quasinormal modes.  Therefore, our poles have not been shown
conclusively to be quasiparticles.  However, in
ref.~\cite{Albash:2009wz}, the pole studied there was the direct
perturbation of the quasiparticle pole in ref.~\cite{Liu:2009dm} under
a magnetic field (meaning that zero magnetic field limit matched) and
the perturbed pole occurred at non--zero $\omega$.  In that case, since
the $H = 0$ pole was a quasiparticle, it followed that the $H \neq
0$ pole would also be a quasiparticle.  Therefore, in the analysis
that is presented next, the fact that the poles occur at non--zero
$\omega$ is not entirely surprising, and they have a chance of being
real quasiparticle excitations.
%
\section{ Numerical Results for the Quasiparticle Spectrum}
\label{sec:numerical_results}
%
We employ the numerical methods that we presented in our earlier
paper\cite{Albash:2009wz} to seek solutions and explicitly construct
the spectral function $G_R(\omega, k)$, using the more general boundary
conditions we have discussed in section~\ref{sec:non-separable} for
the infinite--sum class of solutions. We refer the reader there for
the details of our methods, and here present our observations.

\begin{figure}[ht] 
   \centering
   \includegraphics[width=2.5in]{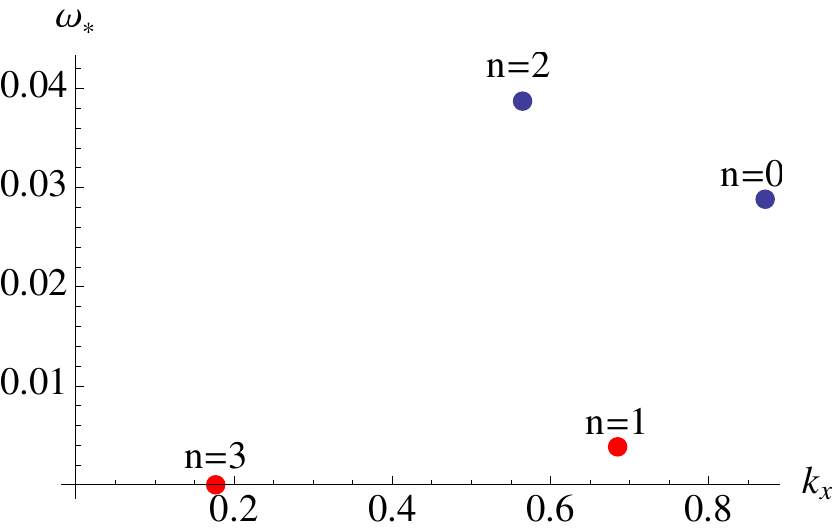} 
   \caption{\small The first (finite) class of peaks for a specific
     value of magnetic field set by $H= -0.1$. }
   \label{fig:H=0_1_first}
\end{figure}

\begin{figure}[ht]
\begin{center}
\subfigure[]{\includegraphics[width=2.5in]{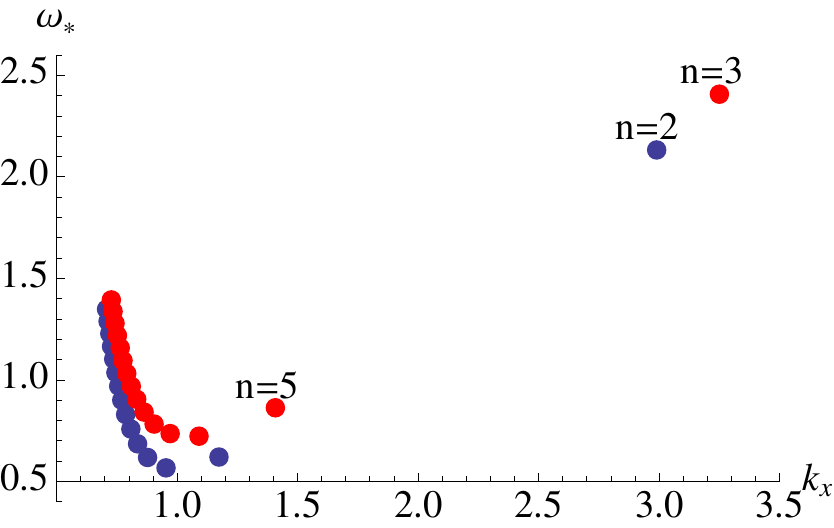}\label{fig:w_vs_k}} \hspace{0.5cm}
\subfigure[$$]{\includegraphics[width=2.5in]{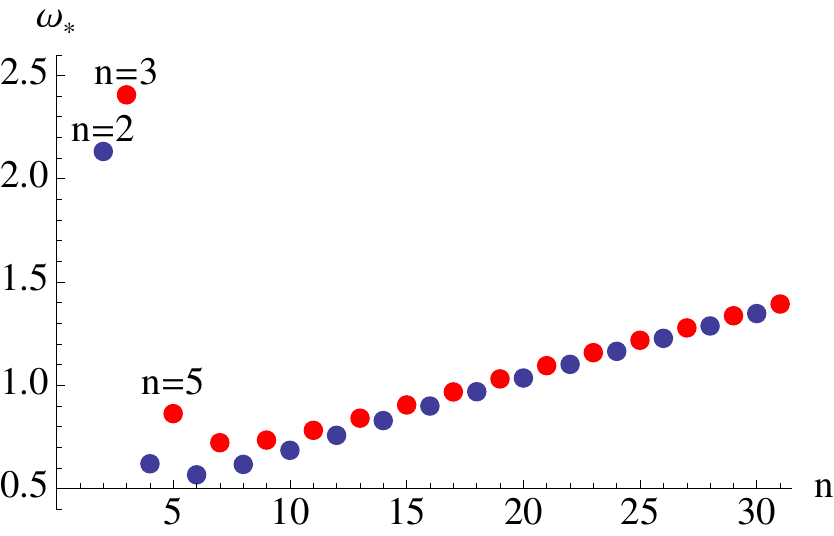}\label{fig:w_vs_n}} 
\caption{\small The second class of peaks for a specific value of magnetic field set by $H= -0.1$. (a) The
  dependence on Fermi momentum. (b) The dependence on Landau level
  number.}  \label{fig:H=0_1_second}
   \end{center}
\end{figure}

First, we notice that for the first four $n$, ($n=0,\ldots,3$) we find
two peaks (signalling the presence of poles in the spectral function)
for a given~$n$. We observed the pair at $n=0$ in our earlier
paper\cite{Albash:2009wz}, and this structure persists for a few
levels before disappearing, leaving only single poles for every $n$
beyond $n=3$.  In each case where there is a pair, the poles fall into
two classes. One class appears to be a deformation of the prototype
pole found in ref.~\cite{Liu:2009dm}. This type is found at smaller
$\omega$. The other, found at larger $\omega$ and $k_x$, appears to be
in the same class as the lone poles present at large $n$, fitting into
a smooth curve of progression.  We present the first class of poles in
figure~\ref{fig:H=0_1_first} and the second class of poles in
figure~\ref{fig:H=0_1_second}. In all figures, $\omega_*$ is used to
denote the value of $\omega$ at which the peak is located. (In both
figures we colour code even $n$ poles with blue and odd $n$ with red.)
Notice that for the first class of poles the even $n$ poles are at
higher $\omega$ while for the second class of poles it is the other
way around.
The behavior of the first poles (at $n=0$) was studied extensively in
ref.~\cite{Albash:2009wz} as a function of $H$, and we expect similar
behaviour here for $n=1,2,3$. Therefore we focus on the second class
of poles, which persists for all Landau levels $n$, in what follows.
We present some results for fixed $H$ and varying $n$ in figure
\ref{fig:H=0_1_second}.  (Recall that the physical magnetic field is
${\cal H}=-2\alpha^2 H$.) We note that the behavior of the poles in
$\omega_*$ {\it vs.} $n$ becomes more predictable at higher
$n$. 
In addition, we note that for $H q > 0$, we find that the solutions
with ansatz~2 (associated with the odd $n$s) have a larger
$\omega_\ast$ than the solutions with ansatz~1 (associated with the
even $n$s).

\begin{figure}[h!]
\begin{center}
\subfigure[$n=10$]{\includegraphics[width=2.25in]{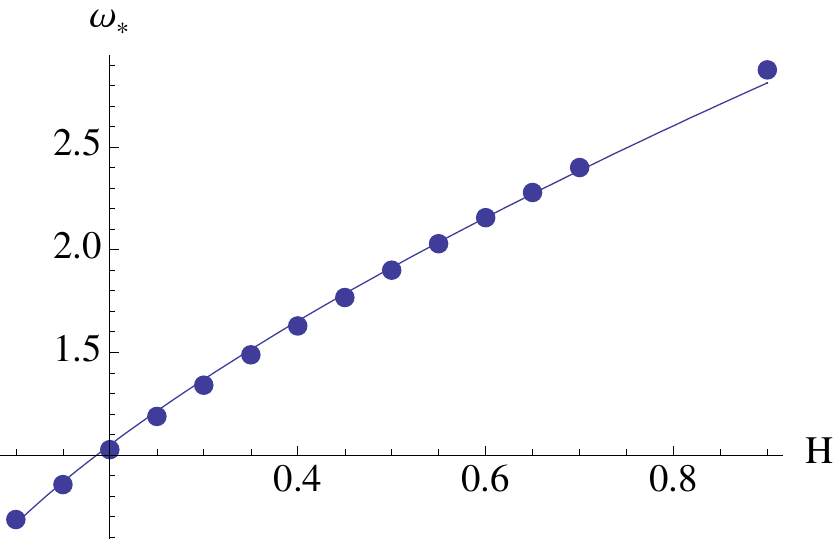}\label{fig:w_vs_H_n=10}} \hspace{0.5cm}
\subfigure[$n=20$]{\includegraphics[width=2.25in]{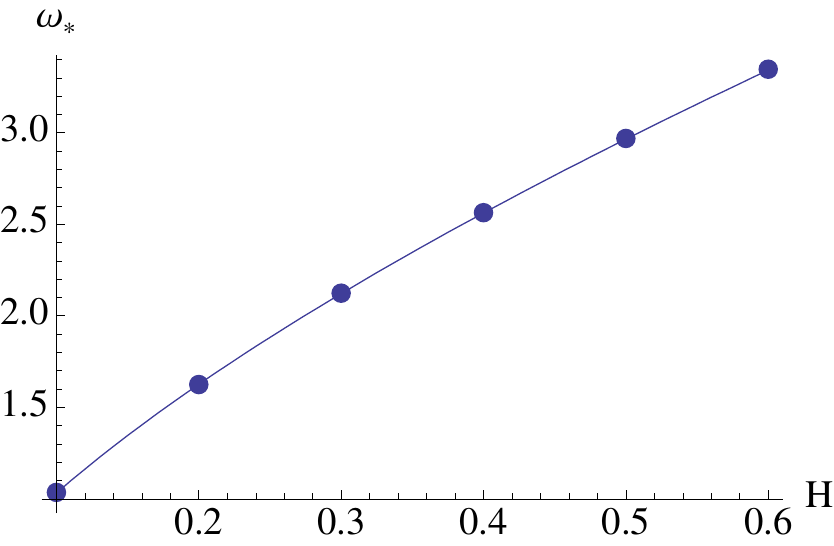}\label{fig:w_vs_H_n=20}} 
\subfigure[$n=30$]{\includegraphics[width=2.25in]{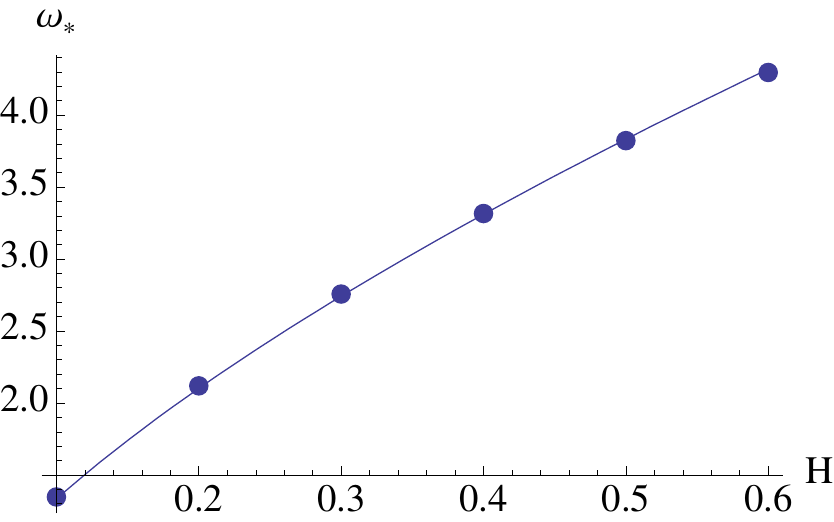}\label{fig:w_vs_H_n=30}} 
\caption{\small Behavior with increasing $H$ for the poles in the regular regime.  The best fit line is given by equation \reef{eqt:bestfit}.}    \label{fig:w_vs_H}
\end{center}
\end{figure}

In figure \ref{fig:w_vs_H} we plot the behavior of the poles at fixed
$n$ for varying $H$, for a few sample values of $n$, choosing of $n$
such that it falls in the regime where $\omega_*$ behaves regularly
with~$n$.
Using these (and much more) data, we can attempt to fit the 
behavior of $\omega_\ast$ with the magnetic field and $n$, for large $n$, giving:
\begin{equation} \label{eqt:bestfit}
\omega_\ast \sim 0.71\,\, H^{0.65} n^{0.63}\ .
\end{equation}
For how well this function fits the data, see figure \ref{fig:w_vs_H}.
In particular, we find that the fit improves for larger $H$ and larger
$n$.  The exact numbers associated with the scaling of $H$ and $n$ are
not as important as the fact that they deviate from the free
relativistic fermion behavior which would have both scaling with
exponent $1/2$.

\begin{figure}[h!]
\begin{center}
\subfigure[$n=0$]{\includegraphics[width=2.25in]{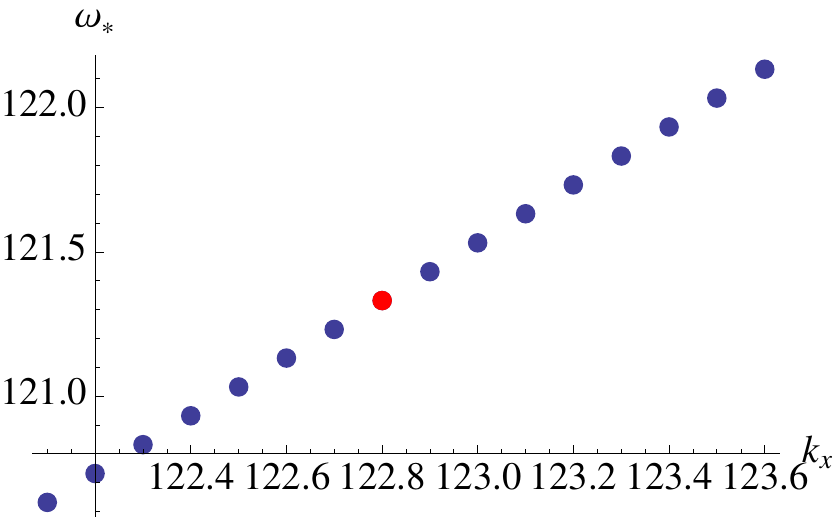}\label{fig:n=0}} \hspace{0.5cm}
\subfigure[$n=1$]{\includegraphics[width=2.25in]{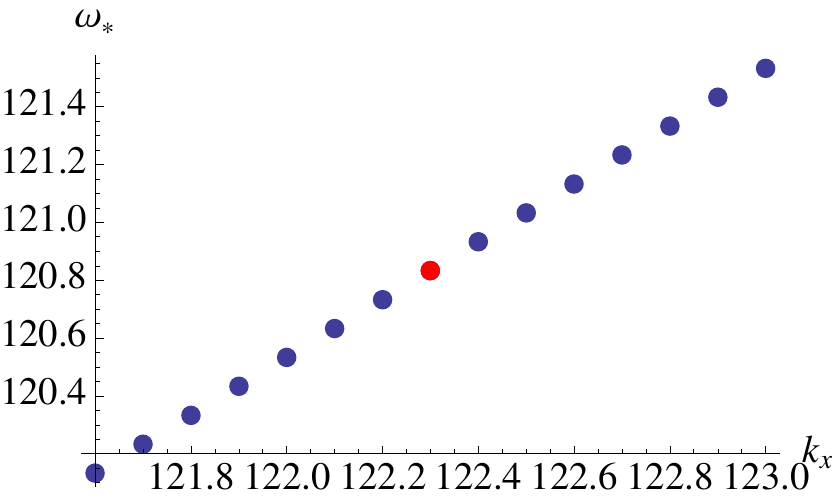}\label{fig:n=1}} 
\subfigure[$n=10$]{\includegraphics[width=2.25in]{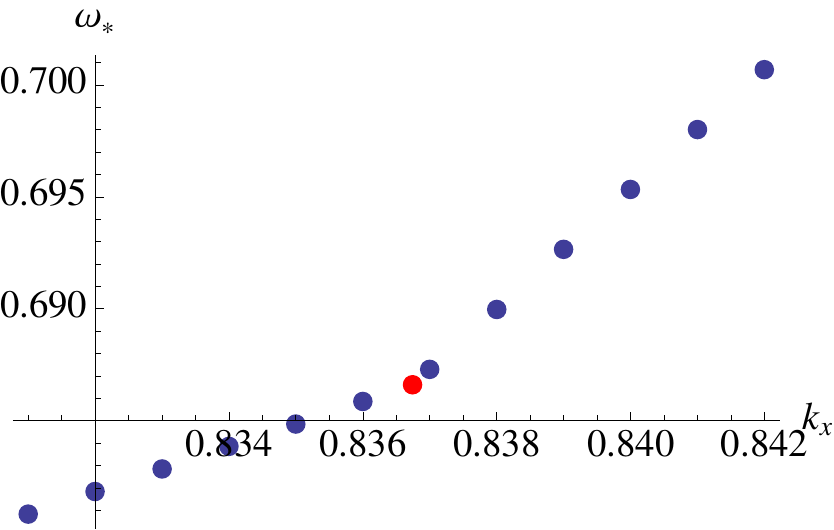}\label{fig:n=10}} \hspace{0.5cm}
\subfigure[$n=11$]{\includegraphics[width=2.25in]{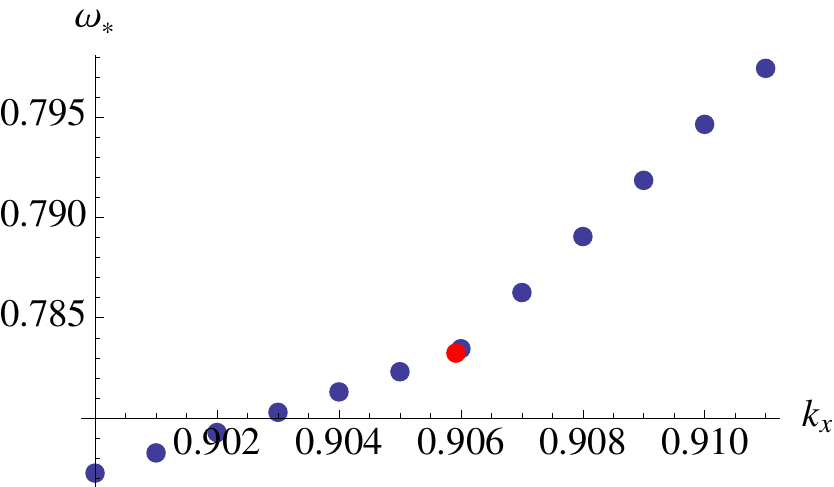}\label{fig:n=11}} 
\subfigure[$n=20$]{\includegraphics[width=2.25in]{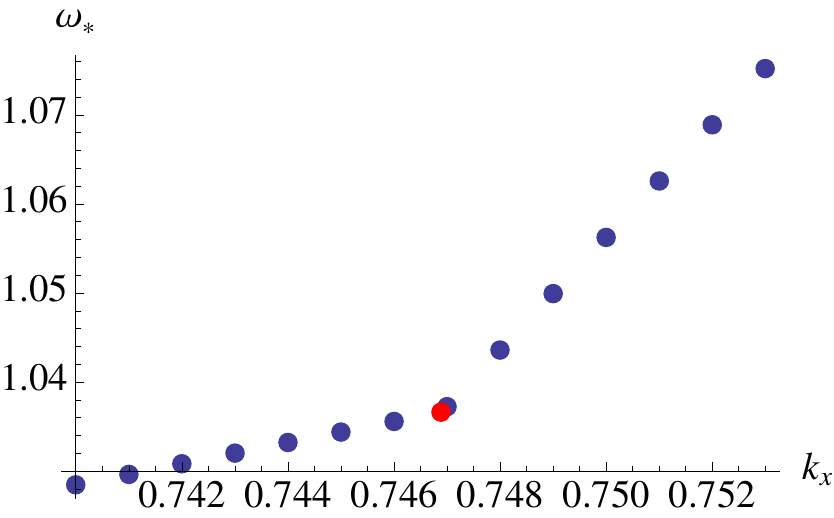}\label{fig:n=20}} \hspace{0.5cm}
\subfigure[$n=21$]{\includegraphics[width=2.25in]{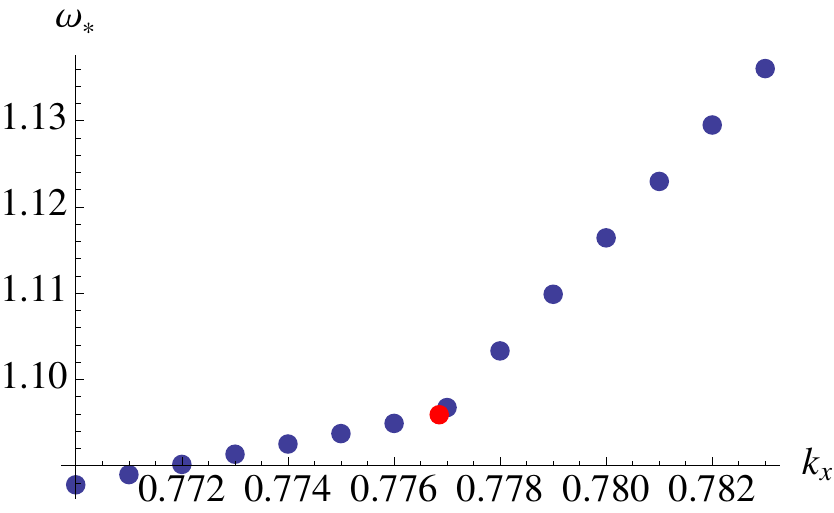}\label{fig:n=21}} 
\subfigure[$n=30$]{\includegraphics[width=2.25in]{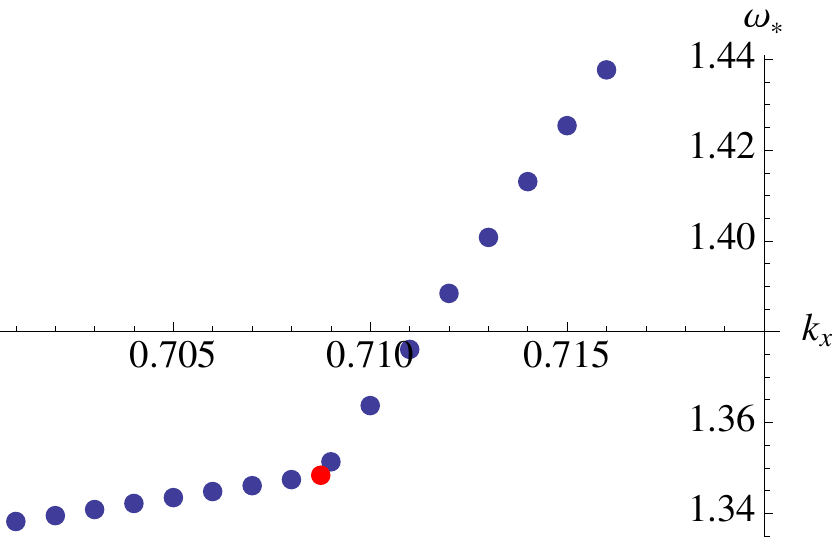}\label{fig:n=30}} \hspace{0.5cm}
\subfigure[$n=31$]{\includegraphics[width=2.25in]{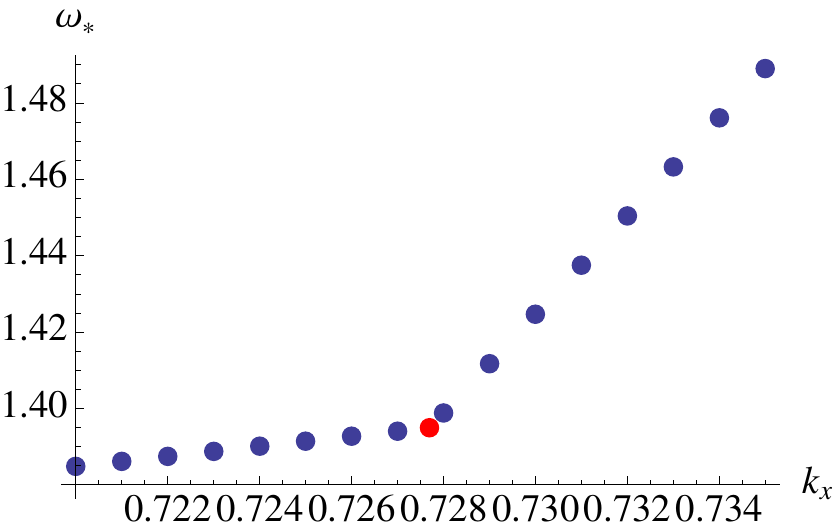}\label{fig:n=31}} 
\caption{\small Dispersion behavior near the pole.  The red dot marks the position of the pole.} \label{fig:dispersion}
\end{center}
\end{figure}

Another important characteristic of these poles is the rate at which
the peak is approached as a function of $k_x$. This is the dispersion
of the quasiparticles associated with that peak, and it can be read
off from our numerical searches quite readily. In our earlier
paper\cite{Albash:2009wz}, we found that non--zero $H$ modified the
behaviour seen in ref.\cite{Liu:2009dm}. In particular, a peak at
higher $k_x$ had more linear ({\it i.e.} Landau--like) dispersion. We
observe that the same is true for all $n$. Note however that from
figure~\ref{fig:H=0_1_second} it can be seen that, for the second
class of peaks (the one that persist for all $n$), at higher $n$, the
peaks are at successively {\it lower} values of $k_x$ (in either the
even or odd $n$ sequence). We show some sample dispersion behavior of
the poles at fixed $H$ but different $n$ in figure
\ref{fig:dispersion}.

\section{Conclusions}
We have generalized the work of ref.~\cite{Albash:2009wz} to include
an infinite number of excitations labeled by $n$. Different $n$
correspond to different Landau Levels for the probe fermion in the
presence of magnetic field.  We studied in detail two separate classes
of solution, leading to distinct physics. One class is separable (it
is used in the work of refs.\cite{Basu:2009qz,Denef:2009yy}) and the
other is an infinite sum of the separable solutions (studied for $n=0$ in our first
paper\cite{Albash:2009wz}).  Both classes are probably physical, but
the first class does not allow a smooth $H=0$ limit for arbitrary~$k_x$, and so seems less suitable for discussion of quasiparticle
peaks and Fermi surfaces in the spirit of ref.\cite{Liu:2009dm}, which
is our focus (and that of ref.\cite{Basu:2009qz}).  In fact, we argued
that the separable solutions have a {\it constant} Green function at
$k_y = 0$ when the $y$--dependence is included in the Green function
definition, which differs from the argument presented in
ref.~\cite{Basu:2009qz}.

The infinite--sum solutions have key properties that make them
attractive.  They limit to the $H=0$ solution smoothly for arbitrary
$n$, with non--trivial dependence on $k_x$.  This is the class of
solutions we use to find quasiparticle peaks for all the Landau
Levels~$n$. Note that since $\omega$ depends on $k_x$, these Landau
levels do not have the usual degeneracy found for the free fermion
Landau levels.

We also noted that levels given by even $n$ and odd $n$ form distinct
towers, distinguished by a relative energy shift (that diminishes as
$n$ increases) that suggests a aligned/anti--aligned coupling to the
magnetic field. The difference between even and odd stems from two
different choices one can make at the event horizon.

We also noticed that the family of quasiparticle peaks, at
sufficiently large $n$, has a dependence on $H$ and $n$ that is
different that of the relativistic massless free fermion.  In fact the
exponent lies between the free relativistic value of $1/2$ and the
non--relativistic value of unity.  This is consistent with our
quasiparticles possibly having an induced mass.

\section*{Acknowledgements}
We would like to thank Nikolay Bobev, Jos\'e Barbon, Karl Landsteiner,
Moshe Rozali, and Julian Sonner for useful conversations.  We would also like to thank an anonymous referee of J. Phys. A. for helpful suggestions for improvement of this manuscript.  This work
was supported by the US Department of Energy.

\providecommand{\href}[2]{#2}\begingroup\raggedright\endgroup

\end{document}